# Efficient 3D Space Time Space Block Code for Future Terrestrial Digital TV

Y. Nasser *member IEEE*, J.-F. Hélard *Senior member IEEE*, M. Crussière and O. Pasquero

*Institute of Electronics and Telecommunications of Rennes, UMR CNRS 6464, Rennes, France*

*Email : youssef.nasser@insa-rennes.fr*

## ABSTRACT

This article introduces a 3D space-time-space block code for future terrestrial digital TV in single frequency networks. The proposed 3D code is based on a double layer structure designed for inter-cell and intra-cell space time coded transmissions. We show that this new structure is particularly efficient for SFN environments whatever the location of the receiver. It is then suitable for fixed, portable and mobile reception.

## 1. INTRODUCTION

Broadcasting digital TV is currently an area of intensive development and standardisation activities. Technically, the terrestrial broadcasting is the most challenging transmission system among the existing digital broadcasting systems due to the presence of strong echoes. The problem becomes even more difficult when broadcasters deploy single frequency networks (SFN) [1] in order to increase the number of TV channels in the allocated frequency bandwidth. SFN are based on the simple addition of lower power transmitters at various sites throughout the coverage area. In an SFN, several transmitters transmit at the same moment the same signal on the same frequency. The overall channel is then modelled as a time-dispersive channel with a long impulse response. Because it is desirable to deploy SFN with lower transmitted powers, increased bit rates and better performance, new multiple input multiple output (MIMO) with orthogonal frequency division multiplexing (OFDM) transmission systems have to be designed to ensure such transmission conditions.

Against this background, a new European CELTIC project called *Broadcast for 21$^{st}$ Century* (B21C) was launched [2]. It constitutes a contribution task force to the reflections engaged by the digital video broadcasting (DVB) project and should give a real support for the conclusions and decisions for a future second generation of terrestrial DVB called DVB-T2. Particularly, it concerns MIMO-OFDM schemes for high definition (HD) TV services.

The work presented in this paper has been carried out within the framework of the B21C project. The contribution of this work is multifold. First, a generalized framework is proposed for modelling the effect of unbalanced powers received from different transmitting antennas. This is a critical problem in SFN with mobile and portable reception. Secondly, we analyze and compare some of the most promising MIMO-OFDM systems in the context of broadcasting for future terrestrial digital TV with equal but also unequal received powers. Eventually, we propose a new 3D space-time-space (STS) block code for SFN environment. The use of a second space dimension is due to SFN. The proposed code is based on the combination of double layer: one layer corresponds to an inter-cell ST coding, the second corresponds to an intra-cell ST coding.

This paper is structured as follows. Section 2 describes SFN architecture. In section 3 we describe the transmission system model. Section 4 gives the receiving model with iterative receiver. In section 5 we discuss the choice of different MIMO schemes considered in this paper and the corresponding simulation results. Section 6 concludes the paper.

## 2. DELAYS AND POWERS IN SFN SYSTEMS

Consider a MIMO-OFDM communication system using $(2 \times M_T)$ transmit antennas (Tx) and $M_R$ receive antennas (Rx) in an SFN environment. Such a system could be implemented on two different sites in an SFN system using $M_T$ Tx by site as shown in Figure 1. Classically, in SFN architectures, the different antennas transmit at the same moment the same signal on the same frequency. Here, we propose to apply a space-time block code (STBC) encoder between different antennas in both sites.

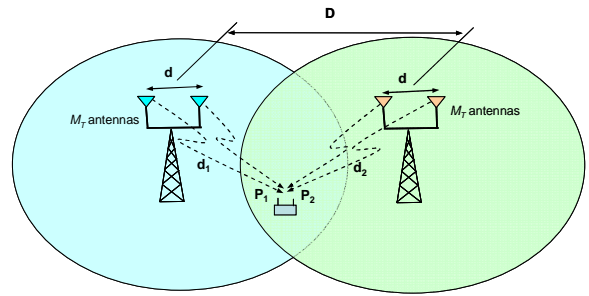

Figure 1- SFN network with unequal received powers

For the SFN to work properly, the time offset between the different received signals must be less than the duration of the guard interval time inserted in the OFDM modulation. As a starting point, let us assume that each site holds one antenna and that the receiver receives signals from both antennas. The time offset between the signals received from each site antennas could be seen as a superposition of the time offset between transmitters' signals (the signal time delay between the transmitting antennas) and the signal time offset between each transmitter and the receiver. The first offset is generally negligible since the transmitters are synchronized with an ultra stable reference like the global positioning system (GPS). The second offset could be seen as follows. When the mobile terminal (MT) moves within one cell, it receives signal from its own cell antenna but also from the neighbouring cell antenna. Since the MT is not equidistant to both antennas, the signal received from each one will be delayed according to the position of the MT.

This results into a delay $\Delta\tau$ between the two received signals from both antennas or equivalently between the channel impulse responses (CIR) between the transmitters and the receiver. The delays are directly related to the distances between the transmitters and the receiver and thus to the signal strength ratio at the receiver. Assuming an equal transmitted power $P_0$ at each antenna, the received power from each antenna is:

$$P_i = \frac{P_0}{d_i^\alpha} \quad (1)$$

where $d_i$ is the distance between the receiver and the $i^{th}$ transmitter and $\alpha$ is the propagation constant which depends on the transmission environment. The delay of each CIR between the $i$th transmitter and the receiver is:

$$\Delta\tau_i = \frac{d_i}{c} \quad (2)$$

where $c_i$ is the light celerity.

Without loss of generality, let us assume that the first transmitter site is the reference site. Substituting $d_i$ from (2) in (1), the CIR delay of the $i$th link (i.e. between the $i$th transmitter and the receiver) with respect to the reference antenna can be expressed by:

$$\Delta\tau_i = \left(10^{\frac{-\beta_i}{10\alpha}} - 1\right)\frac{d_1}{c} \quad (3)$$

where $\beta_i$ is the received power difference (expressed in dB) between the signal received from the reference site and the signal received from the $i$th transmitter. It is given by:

$$\beta_i[dB] = -10 \cdot \alpha \cdot \log_{10}\left(\frac{d_i}{d_1}\right) \quad (4)$$

In the sequel, we will assume that the power received from the reference antenna is equal to 0 dB and the distance $d_i$ is greater than $d_1$ whatever $i$. It is a real situation where the MT is closer to its own cell antenna than other antennas. In this case, $\beta_i$ is neither than the power attenuation factor between the $i$th transmitter and the MT. As a consequence, the transmission model becomes equivalent to a system with unbalanced powers received from each site antennas. If we now consider that $M_T>1$ and $M_R>1$, the choice of an adequate MIMO scheme should then be based on this imbalance in SFN environment and should be adequate for inter-cell and intra-cell environment. This will be the subject of section 4 where we propose a 3D STS code adapted for such situations.

## 3. TRANSMISSION MODEL

Figure 2 depicts the transmitter modules at each site. Information bits $b_k$ are first channel encoded, randomly interleaved, and fed to a quadrature amplitude modulation (QAM) module. The SFN transmission system involving the 2 sites as described in Figure 1 could therefore be seen as a double layer scheme in the space domain. The first layer is seen between the 2 sites separated by $D$ km. The second layer is seen between the antennas separated by $d$ m within one site. For the first layer, a space time block code (STBC) scheme is applied between the 2 signals transmitted by each site antennas. In the second layer, we use a second STBC encoder for each subset of $M_T$ signals transmitted from the same site. For the first layer (respectively the second layer), the STBC encoder takes $L$ (respectively $M$) sets of data complex symbols and transforms them into a $2\times U$ (respectively $M_T\times V$) output matrix according to the STBC scheme. This output is then fed to $2\times M_T$ OFDM modulators, each using $N_c$ subcarriers. The double layer encoding matrix is then described by:

$$\mathbf{X}^{(1)} = \begin{pmatrix} \mathbf{X}_{11}^{(2)} & \cdots & \mathbf{X}_{1U}^{(2)} \\ \mathbf{X}_{21}^{(2)} & \cdots & \mathbf{X}_{2U}^{(2)} \end{pmatrix}$$

$$\mathbf{X}_{pq}^{(2)} = \begin{pmatrix} f_{pq,11}(s_1,...s_M) & \cdots & f_{pq,1V}(s_1,...s_M) \\ \vdots & \ddots & \vdots \\ f_{pq,M_T1}(s_1,...s_M) & \cdots & f_{pq,M_TV}(s_1,...s_M) \end{pmatrix} \quad (5)$$

In (5), the superscript indicates the layer, $f_{pq,it}(s_1,...s_M)$ is a function of the input complex symbols $s_m$ and depends on the STBC encoder scheme. The time dimension of the resulting 3D code is equal to $U\times V$ and the resulting coding rate is $R = \dfrac{L\times M}{U\times V}$.

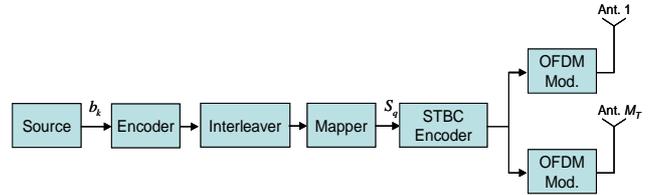

Figure 2- MIMO-OFDM transmitter.

In order to simplify the transmission model, the double layer encoding matrix given in (5) will be represented by $\mathbf{X} = [x_{i,t}]$ where $x_{i,t}$ ($i=1,\ldots,2\times M_T$; $t=1,\ldots,U\times V$) is the output of the double layer STBC encoder on a given subcarrier $n$. In other words, the layers construction is transparent from transmission model viewpoint. Moreover, we set $Q=L\times M$ as the number of the complex symbols at the input of the double layer STBC encoder and we set $T=U\times V$ as the number of the corresponding output symbols. The ST coding rate is then $R =Q/T$. In order to have a fair analysis and comparison between different STBC codes, the signal power at the output of the ST encoder is normalized by $2\times M_T$.

## 4. RECEIVING MODEL

We assume that the transmitter and receiver are perfectly synchronised. Moreover, we assume perfect channel state information (CSI) at the receiver. The signal received on the subcarrier $n$ by the antenna $j$ is a superposition of the transmitted signal by the different antennas multiplied by the channel coefficients $h_{j,i}[n]$ to which white Gaussian noise (WGN) is added. It is given by:

$$y_{j,t}[n] = \sum_{i=1}^{2M_T} \sqrt{P_i}\, h_{j,i}[n] x_{i,t}[n] + w_{j,t}[n] \qquad (6)$$

where $y_{j,t}[n]$ is the signal received on the $n$th subcarrier by the $j$th receiving antenna during the $t$th OFDM symbol duration. $h_{j,i}[n]$ is the frequency channel coefficient assumed to be constant during $T$ symbol durations, $x_{i,t}[n]$ is the signal transmitted by the $i$th antenna and $w_{j,t}[n]$ is the additive WGN with zero mean and variance $N_0/2$. In the sequel, we will drop the subcarrier index $n$ for simplicity. By introducing an equivalent receive matrix $\mathbf{Y} \in \mathbb{C}^{M_R \times T}$ whose elements are the complex received symbols expressed in (6), we can write the received signal on the $n$th subcarrier on all receiving antennas as:

$$\mathbf{Y} = \mathbf{HPX} + \mathbf{W} \qquad (7)$$

where $\mathbf{H}$ is the $(M_R, 2M_T)$ channel matrix whose components are the coefficients $h_{j,i}$, $\mathbf{P}$ is a $(2M_T, 2M_T)$ diagonal matrix containing the signal magnitudes $\sqrt{P_i}$, $\mathbf{X}$ is a $(2M_T, T)$ complex matrix containing the transmitted symbols $x_i[t]$. $\mathbf{W}$ is a $(M_R, T)$ complex matrix corresponding to the WGN.

Let us now describe the transmission link with a general model independently of the ST coding scheme. We separate the real and imaginary parts of the complex symbols input vector $\mathbf{s}$ $\{s_q: q=1,\ldots,Q\}$, of the outputs $\mathbf{X}$ of the double layer ST encoder as well as those of the channel matrix $\mathbf{H}$, and the received signal $\mathbf{Y}$. Let $s_{q,R}$ and $s_{q,I}$ be the real and imaginary parts of $s_q$. The main parameters of the double code are given by its dispersion matrices $\mathbf{U_q}$ and $\mathbf{V_q}$ corresponding (not equal) to the real and imaginary parts of $\mathbf{X}$ respectively. With these notations, $\mathbf{X}$ is given by:

$$\mathbf{X} = \sum_{q=1}^{Q} \left( s_{q,\Re}\mathbf{U_q} + j s_{q,\Im}\mathbf{V_q} \right) \qquad (8)$$

We separate the real and imaginary parts of $\mathbf{S}$, $\mathbf{Y}$ and $\mathbf{X}$ and stack them row-wise in vectors of dimensions $(2Q,1)$, $(2M_RT,1)$ and $(4M_TT,1)$ respectively. We obtain:

$$\begin{aligned}
\mathbf{s} &= \left[ s_{1,\Re}, s_{1,\Im}, \ldots, s_{Q,\Re}, s_{Q,\Im} \right]^{tr} \\
\mathbf{y} &= \left[ y_{1,\Re}, y_{1,\Im}, \ldots, y_{T,\Re}, y_{T,\Im}, \ldots, y_{M_RT,\Re}, y_{M_RT,\Im} \right]^{tr} \\
\mathbf{x} &= \left[ x_{(1,1),\Re}, x_{(1,1),\Im}, \ldots, x_{(2M_T,T),\Re}, x_{(2M_T,T),\Im} \right]^{tr}
\end{aligned} \qquad (9)$$

where $tr$ holds for matrix transpose.

Since, we use linear ST coding, the vector $\mathbf{x}$ can be written as:

$$\mathbf{x} = \mathbf{F}.\mathbf{s} \qquad (10)$$

where $\mathbf{F}$ has the dimensions $(4M_TT, 2Q)$ and is obtained through the dispersion matrices of the real and imaginary parts of $\mathbf{X}$ [3][4]. As we change the formulation of $\mathbf{s}$, $\mathbf{x}$, and $\mathbf{y}$ in (9), it can be shown that vectors $\mathbf{x}$ and $\mathbf{y}$ are related through the matrix $\mathbf{G}$ of dimensions $(2M_RT, 4M_TT)$ such that:

$$\mathbf{y} = \mathbf{GQx} + \mathbf{w} \qquad (11)$$

The matrix $\mathbf{Q}$ is a $(4M_TT, 4M_TT)$ diagonal matrix whose components are given by:

$$\begin{aligned}
B_{i,i} &= \sqrt{P_i} \quad 2.T(p-1)+1 \leq i \leq 2T.p \\
p &= 1,\ldots,2\cdot M_T
\end{aligned} \qquad (12)$$

Matrix $\mathbf{G}$ is composed of blocks $\mathbf{G_{j,i}}$ ($j=1,\ldots,M_R$; $i=1,\ldots,2.M_T$) each having $(2T,2T)$ elements given by:

$$\mathbf{G_{j,i}} = \begin{pmatrix}
h_{(j,i),\Re} & -h_{(j,i),\Im} & 0 & & \cdots & & 0 \\
h_{(j,i),\Im} & h_{(j,i),\Re} & 0 & & \cdots & & 0 \\
0 & 0 & h_{(j,i),\Re} & -h_{(j,i),\Im} & 0 & \cdots & 0 \\
0 & 0 & h_{(j,i),\Im} & h_{(j,i),\Re} & 0 & \cdots & 0 \\
0 & & \cdots & 0 & \ddots & 0 & 0 \\
0 & & \cdots & 0 & \ddots & 0 & 0 \\
0 & & \cdots & & 0 & h_{(j,i),\Re} & -h_{(j,i),\Im} \\
0 & & \cdots & & 0 & h_{(j,i),\Im} & h_{(j,i),\Re}
\end{pmatrix}_{(2T,2T)} \qquad (13)$$

Now, substituting $\mathbf{x}$ from (10) in (11), the relation between $\mathbf{y}$ and $\mathbf{s}$ becomes:

$$\mathbf{y} = \mathbf{GQFs} + \mathbf{w} = \mathbf{G_{eq}s} + \mathbf{w} \qquad (14)$$

$\mathbf{G_{eq}}$ is the equivalent channel matrix between $\mathbf{s}$ and $\mathbf{y}$. It is assumed to be known perfectly at the receiving side.

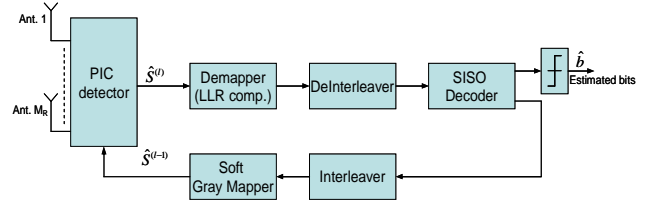

Figure 3- Iterative receiver structure

Now, the detection problem is to find the transmitted data $\mathbf{s}$ given the vector $\mathbf{y}$. The optimal receiver is based on joint ST detection and channel decoding operations. However such receiver is extremely complex to implement and requires large memory for non-orthogonal (NO) STBC codes. Thus the sub-optimal solution proposed here consists of an iterative receiver where the ST detector and channel decoder exchange extrinsic information in an iterative way until the algorithm converges. The iterative detector shown in Figure 3 is composed of a MIMO equalizer, a demapper which is made up of a parallel interference canceller (PIC), a log likelihood ratio (LLR) computation [5], a soft-input soft-output (SISO) decoder [6], and a soft mapper. At the first iteration, the demapper takes the estimated symbols $\hat{\mathbf{s}}$, the knowledge of the channel $\mathbf{G_{eq}}$ and the noise variance, and computes the LLR values of each of the coded bits transmitted per channel use. The estimated symbols $\hat{\mathbf{s}}$ are obtained via minimum mean square error (MMSE) filtering according to:

$$\hat{s}_p = \mathbf{g_p^{tr}} \left( \mathbf{G_{eq}} \cdot \mathbf{G_{eq}^{tr}} + \sigma_w^2 \mathbf{I} \right)^{-1} \mathbf{y} \qquad (15)$$

where $g_p^{tr}$ of dimension ($2M_RT$, 1) is the $p^{th}$ column of $G_{eq}$ ($1 \leq p \leq 2Q$). $\hat{s}_p$ is the estimation of the real part ($p$ odd) or imaginary part ($p$ even) of $s_q$ ($1 \leq q \leq Q$). Once the estimation of the different symbols $s_q$ is achieved by the soft mapper at the first iteration, we use this estimation for the next iterations process. From the second iteration, we perform PIC operation followed by a simple inverse filtering (instead of MMSE filtering at the first iteration):

$$\hat{\mathbf{y}}_\mathbf{p} = \mathbf{y} - \mathbf{G}_{\mathbf{eq,p}} \tilde{\mathbf{s}}_\mathbf{p}$$
$$\hat{s}_p = \frac{1}{\mathbf{g}_\mathbf{p}^{tr} \mathbf{g}_\mathbf{p}} \mathbf{g}_\mathbf{p}^{tr} \hat{\mathbf{y}}_\mathbf{p} \quad (16)$$

where $\mathbf{G}_{\mathbf{eq,p}}$ of dimension ($2M_RT$, $2Q$-1) is the matrix $\mathbf{G}_{\mathbf{eq}}$ with its $p^{th}$ column removed, $\tilde{\mathbf{s}}_\mathbf{p}$ of dimension ($2Q$-1, 1) is the vector $\tilde{\mathbf{s}}$ estimated by the soft mapper with its $p^{th}$ entry removed.

## 5. CHOICE OF MIMO SCHEMES

The aim of this section is to judiciously build the proposed double layer 3D STS code so that the resulting MIMO scheme behaves efficiently in an SFN context. We then need to choose the adequate ST coding scheme to apply on each layer of our 3D code. In the sequel, we consider the orthogonal Alamouti code [7], the space multiplexing (SM) scheme [8] and the Golden code [9]. The first scheme is considered for its robustness, the second is a full rank and the third one is a full rank full diversity code.

### 5.1. Single Layer case: inter-cell ST coding

In the case of single layer reception i.e. $M_T$=1, the second layer matrix $\mathbf{X}^{(2)}$ in (5) resumes to one element. The MIMO transmission is therefore achieved by the set of one antenna in each site. Due to the mobility, the MT is assumed to occupy different locations and the first layer ST scheme must be efficient face to unequal received powers. For equal received powers, we assume that the powers of matrix $\mathbf{Q}$ in (11) are equal to 0 dB. To identify the most efficient ST code under that conditions, simulations have been run under with the parameters of a DVB-T system (see Table 1). The spectral efficiencies 4 and 6 [b/s/Hz] are obtained for different ST schemes as shown in Table 2. Figure 4 gives the required $E_b/N_0$ to obtain a bit error rate (BER) equal to $10^{-4}$ for a spectral efficiency $\eta$=4 [b/s/Hz]. In this figure, we assume that the transmission from a transmitting antenna $i$ to a receiving antenna $j$ is achieved for each subcarrier $n$ through a frequency non-selective Rayleigh fading channel. Moreover, since we have one antenna by site, we set $\beta_1$=0 dB and we change $\beta$= $\beta_2$. As expected, this figure shows that the Golden code presents the best performance when the Rx receives the same power from both sites (i.e. $\beta_1$=$\beta_2$=0 dB). When $\beta_2$ decreases, Alamouti scheme is very efficient and presents only 3 dB loss in terms of required $E_b/N_0$ with respect to equal received powers case. Indeed, for very small values of $\beta$, the transmission scenario becomes equivalent to a transmission scenario with one transmitting antenna.

Table 1- Simulations Parameters

| Number of subcarriers | 8K mode |
|---|---|
| Rate $R_c$ of convolutional code | 1/2, 2/3, 3/4 |
| Polynomial code generator | $(133,171)_o$ |
| Channel estimation | perfect |
| Constellation | 16-QAM, 64-QAM, 256-QAM |
| Spectral Efficiencies | $\eta$= 4 and 6 [b/s/Hz] |

Table 2- Different MIMO schemes and efficiencies

| Spectral Efficiency | ST scheme | ST rate $R$ | Constellation | $R_c$ |
|---|---|---|---|---|
| $\eta$=4 [bit/Sec/Hz] | Alamouti | 1 | 64-QAM | 2/3 |
| | SM | 2 | 16-QAM | 1/2 |
| | Golden | 2 | 16-QAM | 1/2 |
| | 3D code | 2 | 16-QAM | 1/2 |
| $\eta$=6 [bit/Sec/Hz] | Alamouti | 1 | 256-QAM | 3/4 |
| | SM | 2 | 64-QAM | 1/2 |
| | Golden | 2 | 64-QAM | 1/2 |
| | 3D code | 2 | 64-QAM | 1/2 |

### 5.2. Double Layer case

Considering the whole double layer space domain construction, two ST coding schemes have to be assigned to each layer of the proposed system. In this paper, we restrict our study to $M_T$ =2 Tx by site. We propose to construct the first layer with Alamouti scheme, since it is the most resistant for unequal received powers case. In a complementary way, we propose to construct the second layer with the Golden code since it offers the best results in the case of equal received powers. After combination of the 2 space layers with time dimension, (5) yields:

$$X = \frac{1}{\sqrt{5}} \begin{pmatrix} \alpha(s_1+\theta s_2) & \alpha(s_3+\theta s_4) & \alpha(s_5+\theta s_6) & \alpha(s_7+\theta s_8) \\ j\alpha(s_3+\bar{\theta}s_4) & \bar{\alpha}(s_1+\bar{\theta}s_2) & j\bar{\alpha}(s_7+\bar{\theta}s_8) & \bar{\alpha}(s_5+\bar{\theta}s_6) \\ -\alpha^*(s_5^*+\theta^*s_6^*) & -\alpha^*(s_7^*+\theta^*s_8^*) & \alpha^*(s_1^*+\theta^*s_2^*) & \alpha^*(s_3^*+\theta^*s_4^*) \\ j\bar{\alpha}^*(s_7^*+\bar{\theta}^*s_8^*) & -\bar{\alpha}^*(s_5^*+\bar{\theta}^*s_6^*) & -j\bar{\alpha}^*(s_3^*+\bar{\theta}^*s_4^*) & \bar{\alpha}^*(s_1^*+\bar{\theta}^*s_2^*) \end{pmatrix} \quad (17)$$

where $\theta = \frac{1+\sqrt{5}}{2}, \bar{\theta} = 1-\theta, \alpha = 1+j(1-\theta), \bar{\alpha} = 1+j(1-\bar{\theta})$.

Since the distance $d$ between the transmitting antennas in one site is negligible with respect to the distance $D$ (Figure 1), the power attenuation factors in the case of our 3D code are such that $\beta_1$= $\beta_2$= 0dB and $\beta$= $\beta_3$= $\beta_4$. Figure 5 shows the results in terms of required $E_b/N_0$ to obtain a BER equal to $10^{-4}$ for different values of $\beta$ and 3 STBC schemes i.e. our proposed 3D code scheme, the single layer Alamouti and the Golden code schemes assuming Rayleigh i.i.d frequency channel coefficients. In this figure, the value $\beta$ corresponds to $\beta_2$ for the single layer case and to $\beta$ =$\beta_3$ =$\beta_4$ for our 3D code. Figure 5 shows that the proposed scheme presents the best performance whatever the spectral efficiency and the factor $\beta$. Indeed, it is optimized for SFN systems and unbalanced received powers. For $\beta$=-12 dB, the proposed 3D code offers a gain equal to 1.8 dB (respectively 3 dB) with respect to the Alamouti scheme for a spectral efficiency η=4 [b/s/Hz] (resp. η=6 [b/s/Hz]). This gain is greater when it is compared to the Golden code. Moreover, the maximum loss of our code due to unbalanced received powers is equal to 3 dB in terms of $E_b/N_0$. This means that it leads to a powerful code for SFN systems. Figure 6 gives the same kind of results with a MIMO COST207 TU-6 channel model [10]. We assume in this case that the MT is moving with a velocity of 10 km/h and the distance $d_1$ of the reference antenna is equal to 5 km. The CIRs between different transmitters and the MT are delayed according to (3). Once again, The results highlight the superiority of the proposed 3D code in real channel models. The gain is of about 1.5 dB for a spectral efficiency η=4 [b/s/Hz] and 3.1 dB for a spectral efficiency η=6 [b/s/Hz].

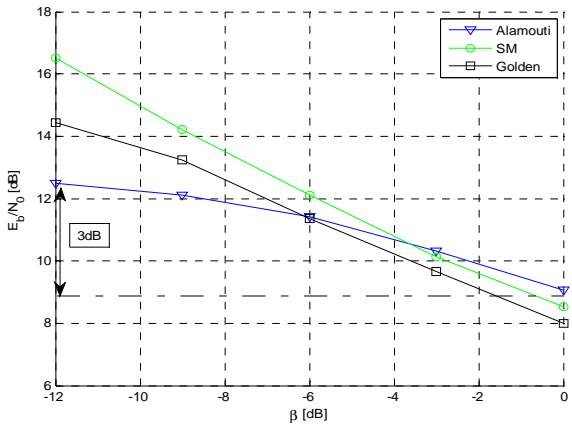

Figure 4- Required Eb/N0 to obtain a BER=$10^{-4}$, single layer case, η=4 [b/s/Hz]

## 6. CONCLUSION

In this paper, a new 3D STSBC is presented. It is based on a double layer structure defined for inter-cell and intra-cell situations by adequately combining the Alamouti code and the Golden code performance. We showed that our proposed scheme is very efficient to cope with equal and unequal received powers in SFN scenarios whatever the receiver position.

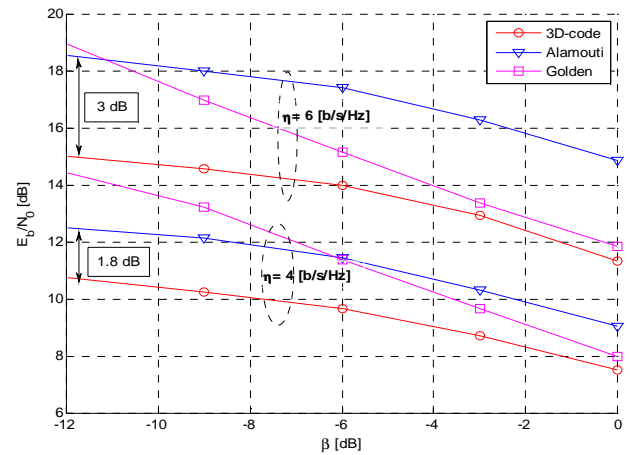

Figure 5- Required Eb/N0 to obtain a BER=$10^{-4}$, double layer case, η=4 [b/s/Hz], η=6 [b/s/Hz], Rayleigh channel

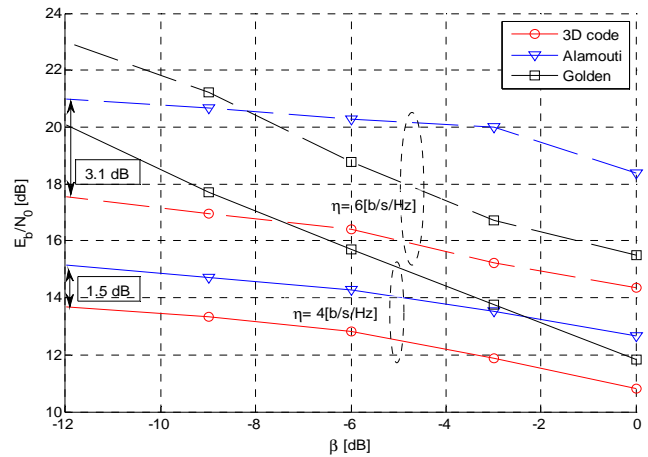

Figure 6- Required Eb/N0 to obtain a BER=$10^{-4}$, double layer case, η=4 [b/s/Hz], η=6 [b/s/Hz], TU-6 channel.